\documentclass[twocolumn,prl,superscriptaddress,showpacs,preprintnumbers,amsmath,amssymb,floatfix]{revtex4}

\usepackage{graphicx}

\begin{document}

\title{Enhancement of vortex pinning in superconductor/ferromagnet bilayers
via angled demagnetization}
\author{Marta Z. Cieplak}
\affiliation{Institute of Physics, Polish Academy of Sciences, 02 668 Warsaw, Poland}
\affiliation{Department of Physics and Astronomy, The Johns Hopkins University,
Baltimore, Md 21218, USA}
\author{L. Y. Zhu}
\altaffiliation[Current address: ]{Materials Science Division, Argonne National Laboratory}
\affiliation{Department of Physics and Astronomy, The Johns Hopkins University,
Baltimore, Md 21218, USA}
\author{Z. Adamus}
\affiliation{Institute of Physics, Polish Academy of Sciences, 02 668 Warsaw, Poland}
\affiliation{Laboratoire des Solides Irradies, \'{E}cole Polytechnique, 91128 Palaiseau,
France}
\author{M. Ko\'{n}czykowski}
\affiliation{Laboratoire des Solides Irradies, \'{E}cole Polytechnique, 91128 Palaiseau,
France}
\author{C. L. Chien}
\affiliation{Department of Physics and Astronomy, The Johns Hopkins University,
Baltimore, Md 21218, USA}
\date{\today}

\begin{abstract}
{We use local and global magnetometry measurements to study the influence of
magnetic domain width $w$ on the domain-induced vortex pinning in
superconducting/ferromagnetic bilayers, built of a Nb film and a ferromagnetic
Co/Pt multilayer with perpendicular magnetic anisotropy, with an insulating layer to eliminate proximity effect. The quasi-periodic
domain patterns with different and systematically adjustable width $w$, as
acquired by a special demagnetization procedure, exert tunable vortex pinning
on a superconducting layer.  The largest enhancement of vortex pinning, by a factor of more than 10, occurs when $w \approx 0.31 \mu$m is close to
the magnetic penetration depth.}
\end{abstract}

\pacs{74.25Ha, 74.25Qt, 74.78Db, 74.78Fk}
\maketitle

One of the most important parameters for the application of type II
superconductors is the critical current density, $J_c$. To achieve high $J_c$
it is necessary to pin vortices, which exist in the mixed state. The usual
method for pinning vortices utilizes microscopic defects in the material that
locally suppress superconductivity and trap normal vortex cores. Magnetic pinning (MP) instead relies on the electromagnetic interaction
between the vortex magnetic field and the stray field generated by the magnetic
texture in the vicinity of the superconductor surface \cite{velez,Luks,aladysh}. Since
magnetic pinning acts on a length scale comparable to the penetration depth,
$\lambda$, it is dominant at temperatures close to the superconducting
transition temperature, $T_c$, where pinning by defects becomes ineffective.

A novel magnetic pinning method utilizes a planar
superconductor(S)/ferromagnet(F) bilayer (SFB) separated by a thin insulating
layer, which eliminates proximity effects \cite{Luks,aladysh,velez,bula,bezpy}. As
previously suggested, the magnetic domains in the F layer create pinning
centers for the vortices in the S layer so that tuning the domain structure can
result in tuning $J_c$ \cite{bula,bezpy}. While the principle of magnetic pinning in
the SFB's has been extensively discussed theoretically \cite{Luks,aladysh,bula,bezpy,Milo,erdin,sonin,laiho,kayali,erdin2} and demonstrated experimentally \cite{garcia,zhang,jan,Lange,Cieplak,Cieplak2,feig,vv,vv2,belkin,karapetrov,leyizhu,vv3,iava,vis}, the magnitude of the pinning enhancement thus far reported has been modest with a
factor of no more than 3, and sometimes even suppression of pinning has been described \cite{feig}. Furthermore, it has not been feasible to compare
various magnetic pinning results, since the different SFB bilayers used in
these studies would render such comparison impractical.

In this work we show that it is possible to systematically vary and achieve
much higher enhancement of $J_c$ in planar SFB's with Nb as the S layer and
Co/Pt multilayer with perpendicular magnetic anisotropy (PMA) as the F layer.
The PMA in the F layer assures a direct effect in magnetic pinning that is not
well defined in F layer with in-plane anisotropy.  Using a single SFB, we use a
special demagnetization method to continuously tune the width of the domains,
$w$, in the quasi-periodic stripe domain pattern of the F layer with equal $+ /
-$ domain, and observe the dramatic effect of tuning on vortex pinning using
the global (SQUID) and the local (Hall sensors) magnetometry measurements.
Using a single SFB with an identical S layer, we have quantitatively determined
$J_c$ enhancement as a result of changing domain width.  We have observed $J_c$
enhancement in excess of 10.

\begin{figure}[t]
\centering
\includegraphics[width=8.5cm]{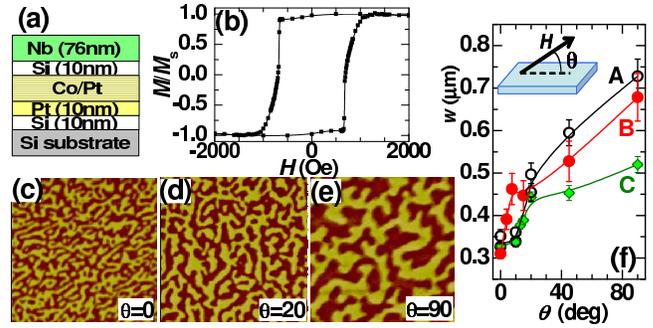}
\caption {(Color online)(a) Schematic of the samples A and B. (b) Hysteresis loop for
sample B at $T = 10$ K. (c-e) MFM images (10 $\mu m$ x 10 $\mu m$) at 300 K for
sample A demagnetized at angles $\theta = 0^{\circ}$, 20$^{\circ}$ and
90$^{\circ}$. (f) $w$ vs $\theta$ for samples A, B and C (lines are guides to
the eye). }\label{fig1}
\end{figure}

The SFB's were grown by sputtering at room temperature on Si(100) substrates
with the same sequence of Si(10)/Pt(10)/[Co(0.6)/Pt(0.3)]$_8$/Si(10)/Nb($t$),
with the thickness denoted in nanometers in parentheses (Fig. 1(a)). The Si(10)
layer between [Co/Pt]$_8$ and Nb eliminates the proximity effect. Two samples
(A and B) with nominally $t$ = 76 nm were studied by magnetometry, whereas the
third sample (C) with $t$ = 20 nm with an additional Si(10) protective layer on
the top was used to extract the vortex activation energy from transport
experiment, as described separately \cite{leyizhu}. All three samples (A, B, C)
display a square hysteresis loop in the normal state, with the coercive fields
at 10 K of $H_c$ = 720 Oe (A), 775 Oe (B), and 750 Oe (C) (Fig. 1(b)). From
magnetoresistance measurements of SFB magnetized to saturation we extract the
superconducting parameters of the Nb layer of $T_c$ = 8.56K (A), 8.42 K (B),
and 7.95 K (C), coherence length, $\xi(0) =$ 14.5 nm (A), 12.4 nm (B), and 12
nm (C), and estimated Ginzburg-Landau parameter for dirty-limit superconductors
\cite{tinkham} of $\kappa \approx $ 3.8 (A), 5.2 (B), and 5.6 (C).

A well known method to acquire multi-domains with equal $+ / -$ domains in an F
layer with PMA is AC demagnetization with field perpendicular to the film
plane.  However, we can also obtain domain pattern with equal $+ / -$ domains
but with different width by demagnetization with an AC magnetic field at an
angle $\theta$ with the sample plane. Figs. 1(c)-(e) show MFM images of sample
A at 300 K for various $\theta$'s. Indeed, the images reveal equal amount of
positive and negative domains, but the average width $w$ increases with
increasing $\theta$. From these MFM images, using two-dimensional Fourier
analysis we extract the mean value of $w$ and the standard deviation, as shown
in Fig. 1(f). Demagnetization at $\theta = 0^{\circ}$ results in a similarly
small $w\approx$ 0.31 - 0.35 $\mu$m for all samples, but the width $w$
increases with $\theta$ with a more rapid growth of $w$ at small $\theta$. This
increase of $w$ with $\theta$ is caused by the fact that in ferromagnets with
PMA the magnetic moments are out-of-plane in the uniform domain area but with
an in-plane component within the domain walls. With increasing $\theta$ the
demagnetization field has a decreasing in-plane component, which tends to align
less spins in-plane and thus creates less domain walls, resulting in a larger
$w$ \cite{leyizhu}. Based on the $T$-dependence of $M_s$ we estimate that the domains may shrink upon cooling down to $T_c$ by less than 10\%.

As alluded to earlier, the samples A and B are nominally the same and yet they
display slightly but noticeably different superconducting transition
temperature $T_c$, coercivity $H_c$, coherence length $\xi(0)$, and the
Ginzburg-Landau parameter $\kappa$.  As shown in Fig. 1, the width $w$ of the
domains for samples A and B is also measurably different. These are unavoidable
sample-to-sample variations for SFB's at these small layer thicknesses, and
underscore the importance of performing experiment with $w$-tuning on a single
SFB sample.

Figure 2(a) shows the hysteresis loops for sample B measured in a SQUID
magnetometer. The loops were measured in a small external magnetic field, $H$,
between -90 Oe and 90 Oe, after setting the domain pattern by AC
demagnetization at angle $\theta$ followed by cooling the sample just below
$T_c$. The cycling of the small magnetic field has no effect on the domain
pattern, and no relaxation of magnetization in the superconducting state has
been observed. Also included in Fig. 2 (a), with the narrowest hysteresis loop,
is the sample with a saturated F layer. The width of the hysteresis loop,
$\Delta M$, increases dramatically as a result of demagnetization indicating a
large enhancement of vortex pinning by the magnetic domains. The largest
$\Delta M$ is at $\theta$ = 0$^{\circ}$, when $w$ is small, and
decreases as $w$ grows with increasing $\theta$. Even at $\theta$ =
90$^{\circ}$ with the largest $w$ of about 0.7 $\mu$m, $\Delta M$
is still much larger than that of the saturated F layer.  We have also observed
a slight asymmetry in the hysteresis curve, particularly at large $\theta$.

\begin{figure}[t]
\centering
\includegraphics[width=8.5cm]{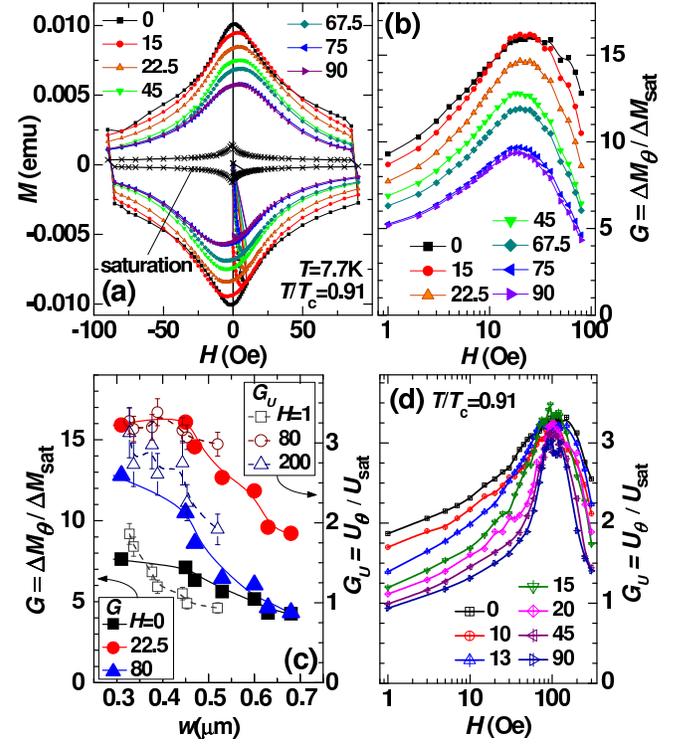}
\caption {(Color online)(a) Hysteresis loops measured with SQUID magnetometer after
demagnetization of sample B with $\theta =$ 0$^{\circ}$, 15$^{\circ}$,
22.5$^{\circ}$, 45$^{\circ}$, 67.5$^{\circ}$, 75$^{\circ}$, 90$^{\circ}$ and
after saturation (from the outermost to the innermost loop, respectively).
Note that loops for 75$^{\circ}$ and 90$^{\circ}$ almost overlap. (b) $G =
\Delta M_\theta / \Delta M_{sat}$ versus $H$ for $\theta =$ 0$^{\circ}$,
15$^{\circ}$, 22.5$^{\circ}$, 45$^{\circ}$, 67.5$^{\circ}$, 75$^{\circ}$,
90$^{\circ}$ (from top to bottom). (c) $G = \Delta M_\theta / \Delta M_{sat}$
(left scale) versus $w$ for $H = $ 0 (full squares), $H = 22.5$ Oe (full
circles), and $H = 80$ Oe (full triangles); and $G_U = U_{\theta} /U_{sat}$ for
sample C, from \cite{leyizhu}, (right scale) versus $w$ for $H = $ 1 Oe (open squares), $H = $ 80 Oe
(open circles) and $H = $ 200 Oe (open triangles). (d) $G_U = U_{\theta}
/U_{sat}$ versus $H$, extracted from transport experiment for sample C \cite{leyizhu}, for
$\theta =$ 0$^{\circ}$, 10$^{\circ}$, 13$^{\circ}$, 15$^{\circ}$, 20$^{\circ}$,
45$^{\circ}$ and 90$^{\circ}$ (from top to the bottom). }\label{fig2}
\end{figure}

To estimate the enhancement of $J_c$ from the global magnetometry results
one needs a model of the critical state with a specific dependence of $J_c$ on
the magnetic induction $B$. The simplest is the Bean model, which assumes $J_c$
to be independent of $B$, leading to the well-known prediction of $J_c \sim
\Delta M$ \cite{poole}. Under this model, the ratio $\Delta M_{\theta} /{\Delta
M}_{sat} \equiv G$ reflects the enhancement of $J_c$ induced by the domain
pattern as set by the demagnetization, where $\Delta M_{\theta}$ and ${\Delta
M}_{sat}$ are the hysteresis loop width for sample demagnetized at $\theta$ and
that of the saturated F layer respectively.  The dependence of $G$ on $H$ is
shown in Fig. 2(b) on a semi-logarithmic scale, omitting the same dependence
for negative $H$. The $G(H)$ dependencies are similar for various $\theta$. The
largest value of $G \approx 16$ is observed for small $\theta$ in the
intermediate field range of $H \sim 15 - 30$ Oe.  The value of $G$ at all
fields decreases with increasing of $\theta$, but still retains a high peak
value of $G \approx 9$ at $\theta \approx 90^{\circ}$. Fig. 2(c) shows the
dependence of $G$ (left scale, full data points) on the actual domain width $w$
for three representative field values, at the peak ($H = 22.5$ Oe), low-$H$ ($H
= 0$), and high-$H$ ($H = 80$ Oe). The $G$ is reduced almost linearly with
increasing $w$ at low-$H$ (bottom curve). At the peak (top curve) the $G$
remains large for narrow domains, and starts to decrease rapidly for $w$
exceeding about 0.45 $\mu$m. The high-$H$ behavior (middle curve) is
intermediate, with the weak suppression of $G$ for small $w$, and more rapid
suppression for large $w$.

To obtain model-independent assessment of $J_c$, we need microscopic
measurement of $B$ at various locations.  For this purpose we employ a linear
array of miniature Hall sensors to probe locally the dependence of $B$ on the
distance $x$ from the sample edge. The local magnetic field is defined as
$H_{loc} = B - {\mu}_0 H$, and $J_c$ is obtained from the relation ${\mu}_0 J_c
\approx 2dB/dx$ \cite{brandt}.  From sample A, we cut a 240 $\mu$m wide strip
and placed a line of 10 sensors, each of which $5 \times 5$ $\mu$m$^2$
in size and 20 $\mu$m apart, across the strip. An additional sensor, placed a
few mm outside the sample edge, is used to measure the actual applied
field as shown in Fig. 3(a). The $B(x)$ is registered simultaneously by Hall sensors while $H$ is swept from 0 up to +100 Oe, and from +100 Oe to -100
Oe. The evolution of $H_{loc} (x)$ during the second part of the sweep is shown
in Figs. 3(b)-(c) for $\theta =$ 0$^{\circ}$ and 90$^{\circ}$. As $H$ is swept,
the flux remains trapped inside, so that $H_{loc}$ increases in the sample
center, eventually reaching a maximum of 41 Oe ($\theta=0^{\circ}$) or 27 Oe
($\theta=90^{\circ}$) just after $H$ changes sign to negative, while in the
saturated sample 5.5 Oe is observed. The calculated $J_c$ has a maximum at
approximately the same $H$, reaching about $1.5 \times 10^8$ Am$^{-2}$ for
$\theta=0^{\circ}$, and $8.4 \times 10^7$ Am$^{-2}$ for $\theta = 90^{\circ}$,
while it is only $1.6 \times 10^7$ Am$^{-2}$ in the saturated sample. This
confirms the conclusion of strong pinning enhancement, particularly for
narrow domains at $\theta = 0^{\circ}$. Interestingly, Figs. 3(b)-(c)
reveal that the change of $H_{loc}$ is much smoother during the sweep from +100
Oe to 0 than during the sweep from 0 to -100 Oe, where there are many sudden
and large drops of $H_{loc}$ across the whole sample. Such abrupt decrease of
$H_{loc}$ indicates the abrupt annihilation of flux over the large sample area.
It is most likely triggered by the strong interaction between the positive flux
still trapped inside and negative flux entering the sample. This effect
likely contributes to the asymmetry of hysteresis seen in Fig. 2(a).

\begin{figure}[t]
\centering
\includegraphics[width=8.5cm]{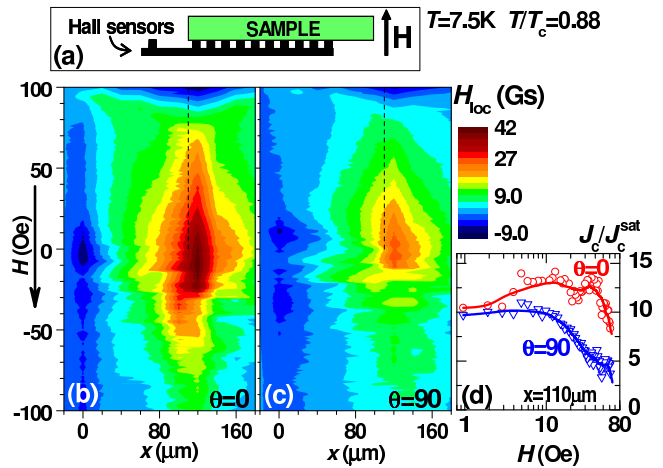}
\caption {(Color online)(a) Placement of the sample, with Nb adjacent to Hall sensors.
(b-c) $H_{loc}(x,H)$ for sample A during the field sweep from +100 Oe to -100
Oe for $\theta = 0^{\circ}$ and $\theta = 90^{\circ}$. The edge and the center
of the sample strip are at $x = 0$, and $x = 120$ $\mu$m, respectively. (d)
$J_c /J_c^{sat}$ vs $H$ calculated for $x = 110$ $\mu$m ($x$ shown
by dashed line in (b-c)). }\label{fig3}
\end{figure}

Figure 3(d) shows the ratio of $J_c$ to $J_c^{sat}$ (where $J_c^{sat}$ is $J_c$
in the saturated sample), calculated for $x = 110 \mu$m, the approximate
position at which $d B /d x$ is the largest (along the dashed line in Figs.
3(b)-(c)). The calculation has been made for $H > 0$, since for negative $H$ $J_c$ is highly irregular because of the jumps of $H_{loc}$. As shown in
Fig. 3(d), $J_c /J_c^{sat}$ is higher for $\theta = 0^{\circ}$ in the whole
$H$-range than that for $\theta = 90^{\circ}$, the same conclusion as that in
Fig. 2(b), although the shape of $J_c /J_c^{sat}$, while generally similar to
$G(H)$ in Fig. 2(b), differs noticeably in details. This is not surprising
because the local measurements using an array of Hall sensors are more
susceptible to local irregularities of the flux penetration inside the narrow
sample, while the global magnetometry measurement integrates over the large
sample. Nevertheless, the most essential conclusion that pinning is strongest for narrow domains remains unchanged. Similar result is obtained
at lower $T$, but the maximal magnitude of the $J_c /J_c^{sat}$ decreases, by
about 17\% at $T/T_c =0.82$, and by 30\% at $T/T_c =0.78$.

The results clearly establish that the strongest pinning occurs at the smallest
$w$. This is due to several factors. The magnitude of the stray field at the
domain center scales as $1/w$ \cite{maksimova}, and moreover, the density of
pinning centers also increases with decreasing $w$. As a result, at low $H$ the
vortices are most effectively pinned when $w$ is the smallest. An additional
argument comes from the comparison of $w$ to the penetration depth, since this
is the length scale of the vortex-domain interaction. The thickness of Nb in
sample B is $t = 76$ nm, slightly larger than $\lambda (0) \approx \kappa \xi
(0) \approx 64$ nm. As $T$ increases, $\lambda (T)$ grows as $\sim \kappa
\xi(0)/\sqrt{1-T/T_c}$, so at $T/T_c =0.91$ we obtain $\lambda \approx 220$ nm.
However, at high $T$ one approaches the limit of $\lambda \gg t$, for which one
should use the effective penetration depth, $\Lambda = \lambda^2 /t \approx
640$ nm \cite{tinkham}. While this higher limit is not quite reached here, it
is likely that $\lambda$ approaches, and perhaps even exceeds, the average
domain width of $w \approx 310$ nm for $\theta = 0^{\circ}$. Since in the
quasi-ordered domain pattern there are areas in which $w$ is smaller than
average, vortices cannot move freely along domains, thus enhancing the pinning.
The value of $w$ grows with increasing $\theta$, and the vortices flow more
easily, thus diminishing the value of $G$.

As shown in Fig. 2(b), for each value of $\theta$, $G$ increases with $H$
reaching a maximum presumably when the vortex lattice becomes commensurate with
the magnetic domain pattern. This is followed by a decrease of $G$, when the
vortex density exceeds that of the domain-induced pinning centers. Since the
domain density decreases with increasing $w$, the suppression of $G$ at high
$H$ is weaker for smaller $w$ as shown by the blue triangles in Fig. 2(c).

The commensurability of vortex pinning can be advantageously observed using an
ordered array of pinning centers, which lead to enhanced pinning at
well-defined matching fields \cite{velez}. In the SFBs the vortices are
confined to the irregular domains of one sign. In the case of regular stripe
domain pattern with the period $2w$ single or multiple chains of vortices are
pinned by each domain, as recently shown by scanning tunneling microscopy
\cite{karapetrov}. We have recently performed transport measurements on sample
C demagnetized at different $\theta$ to capture different domain widths and we
have inferred a similar chain-pinning phenomenon \cite{leyizhu}. At $T$
sufficiently below $T_c$ the resistance is thermally activated, with the
activation energy for vortex pinning, $U$, enhanced by magnetic domains. The
$U(H)$ exhibits maxima at matching fields when narrow domains (small $\theta$)
pin single vortex chains, and wider domains (large $\theta$) pin double vortex
chains.

It is interesting to compare the activation energy $U$ from transport
measurements on sample C \cite{leyizhu} with the present data on sample B. From the transport
results, we calculate the enhancement of $U$ induced by demagnetization at
$T/T_c =0.91$, using the parameter $G_U = U_{\theta} /U_{sat}$, where
$U_{\theta}$ is for sample C demagnetized at $\theta$, and $U_{sat}$ is for
saturated sample. The values of $G_U$ versus $H$ for several values of $\theta$
are shown in Fig. 2(d).  One notes that $G_U$ and $G$
display strikingly similar $H$ dependence. Furthermore, both display the
largest value at the smallest values of $\theta$, decreasing monotonically with
increasing $\theta$. These systematic dependencies are illustrated in Fig.
2(c), where $G$ (filled symbols) and $G_U$ (open symbols)  are for three
representative fields, from the low field, peak, and high field regions ($H$ =
0, 22.5, and 80 Oe for $G$, and $H$ = 1, 80, and 200 Oe for $G_U$,
respectively). The qualitative similarity suggests a common origin, which is
pinning of vortex chains by domains.

However, there are also noticeable differences between $G(H)$ and $G_U(H)$. One
notes that the peak in $G_U(H)$ has a narrower width but it is located at a
higher $H$ than that of $G(H)$, whereas $G$ reaches a maximum magnitude about 5
times larger than that of $G_U$. Most of these differences are due to the
different Nb thickness in sample B (76 nm) for $G(H)$ and sample C (20 nm) for
$G_U(H)$. In addition, sample B shows larger dispersion of $w$. The dispersion
of $w$ contributes directly to the broadening of matching field, while a
thicker Nb layer leads to a progressive reduction of the domain-induced stray
field (by up to about 25\% on the other side of Nb \cite{maksimova}), and therefore reduction of the density of pinning centers across $t$,
shifting $G(H)$ peak to smaller $H$. The difference in magnitude between $G$
and $G_U$ is also mainly related to thickness.  The strength of the magnetic
interaction roughly depends on the ratio of the domain width $w$, to the range
of magnetic interaction, given approximately by $\Lambda$. In the thicker
sample B the magnetic interaction is stronger because $w$ is comparable to
$\Lambda$, whereas in the thinner sample C, $\Lambda$ is about 4 times larger
so that the magnetic interaction is substantially reduced.

In conclusion, we have demonstrated a method to induce large enhancement of
vortex pinning in the superconductor/ferromagnet bilayers using a ferromagnetic
layer with perpendicular magnetic anisotropy. By demagnetizing the sample at an
angle to the sample surface, we obtain domain patterns with equal $+/-$ domains
but different domain width $w$.  This unique attribute allows a single
bilayer to acquire different domain width to
exert tunable vortex pinning on a superconducting layer. Magnetometry and
local measurements using an arrays of Hall sensors show conclusively vortex
pinning enhancement, by a factor of more than 10 at domain width $w$ of about
310 nm, much larger than previously obtained.

This work was supported by Polish MNiSW grant N202 058 32/1202, by NSF grant
DMR05-20491, by the French-Polish Bilateral Program PICS 4916, and by the
European Union within the European Regional Development Fund, through the
Innovative Economy grant POIG.01.01.02-00-108/09.

\end{document}